**Graphical Abstract**

Up today, nanostructure engineering is the only one approach to obtain the large birefringence up to 1.0. Employed first-principles calculations, we demonstrate that $LaOBiS_2$ has a *giant inherent* birefringence (negative), which absolute value is larger than 1.0 and about 4.5 times larger than that of $YVO_4$ at about 600nm wavelength. Hence, the corresponding optical devices can be miniaturized in the scale shown in the Figure, i.e., $LaOBiS_2$:$YVO_4$:$LiNbO_3$ is 1:4.5:12.5 for the same birefringence applications. While the two later crystals are commercially available, $LaOBiS_2$ is highly expected to be experimentally growth and nanostructure engineering.

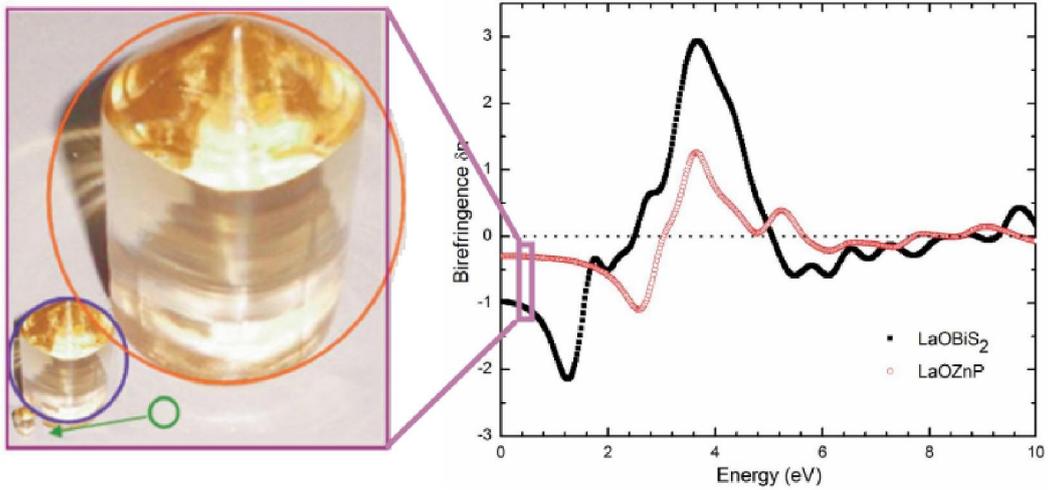

# Giant birefringence in layered compound LaOBiS$_2$

Hai Wang[*]

*College of Materials Science and Engineering, Tongji University, Shanghai 201804, China*

**Abstract**

Birefringent crystals (BFC) have been extensively used in imaging spectrometer, laser devices, and optical components. Seeking new BFC is of important for both fundamental research and industrial application. Employed first-principles density functional theory (DFT), we found that LaOBiS$_2$ (space group P4/nmm), the parent phase of recently discovered BiS$_2$-superconductor, exhibits superior birefringence with a maximum value of 2.94, which is 13.4 times larger than the extensively used YVO$_4$ (0.22). Furthermore, LaOBiS$_2$ is an indirect semiconductor with a band-gap of 0.9 eV, which is suitable for optical applications. The origin of giant inherent birefringence is discussed.

[*] Corresponding author, E-mail addresses: hwang@tongji.edu.cn

Layered compounds have received increasing attentions due to their excellent superconductivity,[1] thermoelectricity,[2] giant magnetoresistance,[3] and so on. They exhibit various electronic, optical and magnetic properties as well as good tunability. For example, the artificial layered-structures are designed for high-efficiency light-emitting diodes (LED)[4] and optical communication[5]. Hosono et al. have explored novel layered oxides $LaT_MPnO$, where a pnicogen anion ($Pn^{3-}$) replaces a chalcogen anion ($Ch^{2-}$) in LaCuOCh, a transition metal cation ($T_M^{2+}$) can substitute for $Cu^+$. There are many combinations ($T_M$ and Pn) to form 2-dimensional (2D) layered structure. Interestingly, the doped-F LaOFeAs is found to be a superconductor with Tc as large as 26 K. The discovery has received the world-wide attentions on iron-based superconductivity. The large number of other layered compounds is examined to reveal their superconductivity. Recently, a novel $BiS_2$-based layered superconductor $LaO_{1-x}F_xBiS_2$ (Tc~10.6K) is report.[6] Its parent compound $LaOBiS_2$ has LaFeAsO-type structure and is a semiconductor rather than a metal as LaOFeAs.[7] More and more research is focused in the superconductivity. As we mentioned above, layered compounds have also various excellent properties. To demonstrate this, used $LaOBiS_2$ as an example, we reveal a giant birefringence based on first-principles density function calculations. We also discussed the origin and proposed the promising potential applications.

The calculations were performed using density functional theory with generalized gradient approximation (GGA) in the scheme of Perdew–Burke–Ernzerhof.[8] The ion-electron interaction was modeled by PAW[9] pseudopotentials in VASP.[10] A plane-wave cutoff energy of 500 eV was employed throughout the calculation. For the sampling of the Brillouin zone, the electronic structures and optical properties used 12x12x6 and 18x18x9 Monkhorst-Pack[11] k-point grids, respectively. Whenever possible, we cross–check the results using Quantum ESPRESSO package (QE).[12] The consistency of our results for two sets of calculations is satisfactory. Employed DFT calculations, we have report the optical of $ZnSnO_3$, where the maximum of birefringence is only 0.2 at 6.90 eV.[13] In the following year, we found the giant optical anisotropy in the parent compounds of iron-based superconductors: LaFeAsO, $BaFe_2As_2$ and LiFeAs.[14]

$LaOBiS_2$ has a layered structure with space group P4/nmm (No. 129), as iron-based superconductor LaFeAsO. Its structure is composed of one $La_2O_2$ and two $BiS_2$ layers (see, Fig. 1). Table I lists our calculated results on the structural parameters of $LaOBiS_2$ as well as the corresponding experimental and theoretical values. Awana *et al.*[7] have report the detailed structural parameters. It is clear that our results are in good agreement with previous reports. Note that only S1 ions have large difference as compared to experimental report, similar theoretical results present in the case of LaFeAsO.

The band structures along high-symmetry are depicted in Fig. 2. It is clear that LaOBiS$_2$ is a semiconductor. The top of valence band locates at Γ (or Z), while the bottom of conductor band at R (or X). Hence, an indirect gap of 0.90 is formed between Γ and R (or Z and X), which is consistent with the VASP value of 0.7 eV [see Figure 5(b)].[7] The GGA band-gap is usually underestimated. Unfortunately, there is no experimental value report so far. The following optical properties are calculated by PBE without scissors correction.

Fig. 3 shows that the optical properties of LaOBiS$_2$ have large anisotropy both in absorptions and refractive index. This is expected due to its layered-structure. For example, the infinite-layer iron oxide SrFeO$_2$ has a giant optical anisotropy.[17] Note that birefringence is the optical property of a material having a refractive index that depends on the polarization and propagation direction of light.[18] It is quantified by the maximum difference in refractive index within the material. This effect was first described in calcite by the Danish scientist Rasmus Bartholin in 1669. There are three well-known natural materials that are birefringent in bulk form: calcite CaCO$_3$ ($\delta n$=0.172), quartz SiO$_2$ ($\delta n$=0.009), and magnesium fluoride MgF$_2$ ($\delta n$=0.006), all measured at a 590nm wavelength.[18] Among artifical crystals, YVO$_4$ has an excellent birefringence of about 0.22, which is inferior to rutile TiO$_2$ (0.28). Owing to its easy growth, large index of refractivity and good mechanical properties, YVO$_4$ crystals have been extensively used in laser devices and optical components.[19] Whatever, seeking new birefringent crystals is of important for both fundamental research and industrial application, especially the device miniaturization.

Now we check the birefringence of LaOBiS$_2$. Surprisingly, we found that its birefringence is large as 2.94 (the maximum), which is 13.4 times larger than that of the extensively used laser crystals YVO$_4$ (0.22). This is *exciting*. As you known that the birefringence of SrFeO$_2$ is also only 0.2-0.5.[17] To the best of our knowledge, the value is the largest among the inherent birefringence so far. In the range of [0, 0.8] eV bellow the band-gap value, the birefringence of LaOBiS$_2$ is negative and its absolute value is larger than 1.0, which also at least 4.5 times larger than that of YVO$_4$, the most important optical materials in optical communication. It is well-known that Nd:YVO$_4$ crystals have been used to generate a radially polarized laser, because YVO$_4$ has more than 3 times larger birefringence than that of LiNbO$_3$ (0.08).[20] Compared to LiNbO$_3$, LaOBiS$_2$ has 12.5 times larger birefringence, which can make device design more compact. In this context, LaOBiS$_2$ may have promising potential applications in semiconductor laser devices, optical communications and so on. For optical display, it can be used to improve grey-scale stability and contrast ratio, i.e., birefringent optical compensators are used to correct the polarization state of light entering and exiting the liquid crystal cell at oblique angles. In addition, owing to its semiconductor and large anisotropic conductivity (not shown here), LaOBiS$_2$ may have very promising potential applications in thermoelectricity and solar cells.

Up today, it seems that nanostructure engineering is only one effective approach to obtain the larger birefringence, which is expected for practical applications. For example, giant birefringence is report in nanostructured silicon thin-film (0.3)[21] and GaP nanowires (0.7).[22] For the later, the birefringence is enhanced to be about 75 times larger than that of natural quartz crystal (0.009). Here, the birefringence of LaOBiS$_2$, we report in this work, is 111 times larger than that of quartz crystal. Hence, its birefringence is *giant*. Recently, the in-plane birefringence of ZnO nanowire arrays is almost one order of magnitude higher than that of quartz.[23] The Glancing Angle Deposition (GLAD)[24] method has been demonstrated to be an effective method to obtain tunable birefringence in a large continuous range. For example, the in-plane birefringence (up to 0.4) of Si thin-films is large compared with naturally anisotropic crystals and is comparable to two-dimensional photonic crystals.[25] In this context, the corresponding experimental investigation on the nanostructure engineering of LaOBiS$_2$ is highly expected. Whatever, the giant birefringence of LaOBiS$_2$ means that it is wonderful idea to study the other properties of layered-structure compounds, rather than only its superconductivity. Here, our results demonstrate DFT is an effective method for this issue.

Why LaOBiS$_2$ have a giant birefringence? To explain the reason, we calculated the birefringence of LaOZnP, a layered compound with the same space group structure (P4/nmm) of LaOBiS$_2$. The results depicted in Fig. 4. While the negative maximum -1.10 is located at 2.57 eV, the positive maximum is 1.26 at 3.64 eV. The corresponding maxima of LaOBiS$_2$ are -2.14 (2.94) at 1.25 eV (3.66 eV), which are about 2 times larger than these of LaOZnP. For LaOZnP, its birefringence is only comparable to YVO$_4$ (0.22). The comparison with LaOZnP indicates that the giant birefringence can not only result from the layered-structures, although the crystals with layered-structures are well-known to be birefringent.

While LaOZnP is 1111-type compound, LaOBiS$_2$ is 1112-type one. This means that S ions in LaOBiS$_2$ are 2 times than P ions in LaOZnP. Compared the structures of both compounds, we find that redundant

S ions are formed Bi-S bonds along z-axis, which enhance the optical anisotropy and birefringence. Why this 1112-type structures can exist，it is not easy to answer. The further investigation is highly expected. Whatever, we conjecture that the lone pair of $Bi^{3+}$ ions may play an important role in this issue.

In summery, we present first-principles calculations on the structural, electronic and optical properties of $LaOBiS_2$. The structural parameters we obtained are in good agreement with previous theoretical values. Band structures reveal that it is an indirect semiconductor with a band-gap of 0.9 eV. There is large optical anisotropy due to its layered-structure. Interestingly, its *inherent* birefringence is found to be the largest among inorganic compounds as we known so far. The giant birefringence is attributed to its layered-structure and $Bi^{3+}$ lone pair. Finally, $LaOBiS_2$ is suggested have a very promising potential applications in laser devices, optical communications, thermoelectricity and solar cells.

Table I. Lattice parameters and Wyckoff positions of LaOBiS$_2$. O sites at origin (0, 0, 0).

|  | Exp.[a] | Theo.[b] | Theo.[c] | This work.[d] | This work.[e] |
|---|---|---|---|---|---|
| $a$ | 4.066 | 4.039 | 4.0677 | 4.043 | 4.051 |
| $c$ | 13.862 | 14.123 | 14.3085 | 14.220 | 14.247 |
| La (0.5,0,z) | 0.090 | 0.0889 | 0.0890 | 0.0855 | 0.0879 |
| Bi (0.5,0,z) | 0.632 | 0.6304 | 0.6309 | 0.6328 | 0.6321 |
| S1 (0.5,0,z) | 0.371 | 0.3932 | 0.3945 | 0.3950 | 0.3937 |
| S2 (0.5,0,z) | 0.819 | 0.8090 | 0.8073 | 0.8111 | 0.8089 |

[a]Experimental, Ref.7. Theoretical, [b]QE, Ref.15. and [c]QE, Ref.16, 60Ry/600Ry. [d]QE. [e]VASP.

**Figure caption**

Fig. 1. Crystal structure of LaOBiS$_2$. Here, Violet stands for Bi, Green La, Yellow S and Red O.

Fig. 2. Band structure of LaOBiS$_2$. Fermi level locates at zero; the band-gap is 0.90 eV.

Fig. 3. Absorption (a), refractive index and birefringence δn (b) of LaOBiS$_2$.

Fig. 4. Birefringence of LaOBiS$_2$ (solid square) and LaOZnP (open circle).

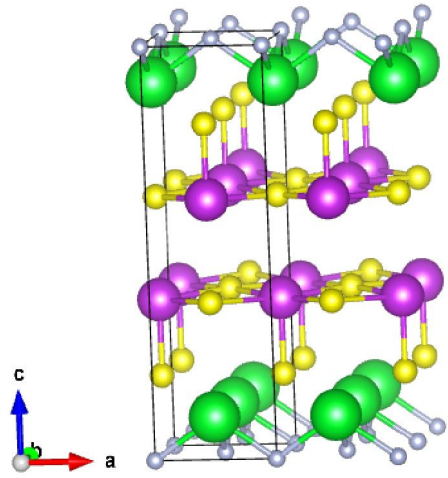

Fig. 1. Crystal structure of LaOBiS$_2$. Here, Violet stands for Bi, Green La, Yellow S and Red O.

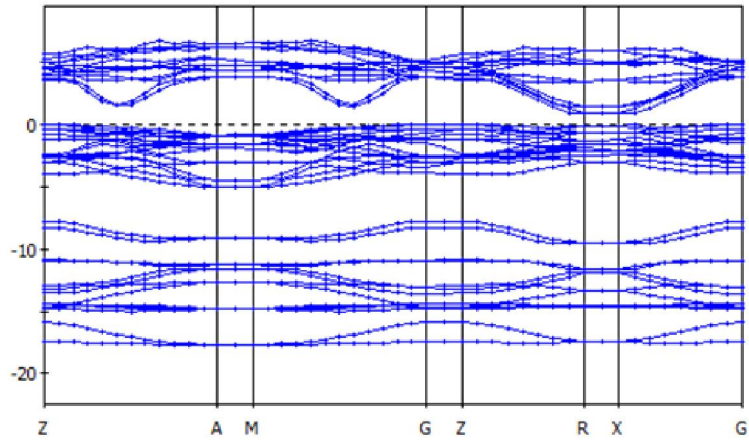

Fig. 2. Band structure of LaOBiS$_2$. Fermi level locates at zero; the band-gap is 0.90 eV.

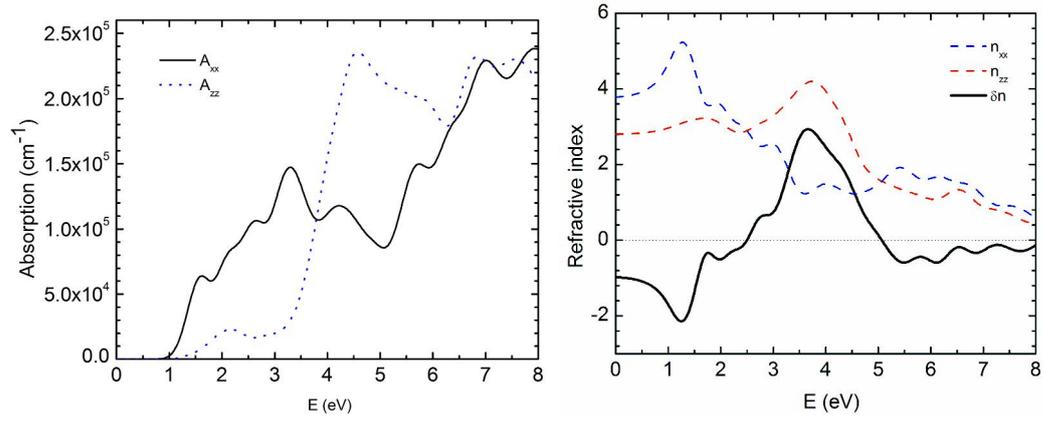

Fig. 3. Absorption (a), refractive index and birefringence δn (b) of LaOBiS$_2$.

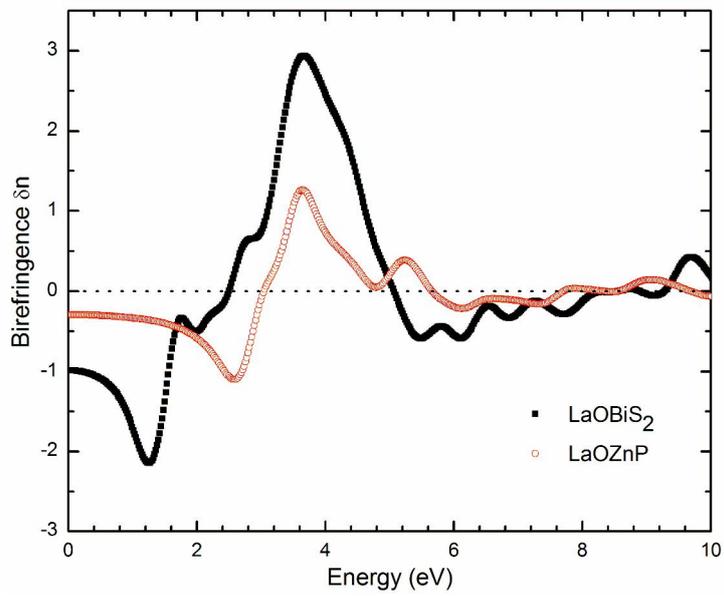

Fig. 4. Birefringence of LaOBiS$_2$ (solid square) and LaOZnP (open circle).